\documentclass[twocolumn,epjc3]{svjour3}          

\RequirePackage[T1]{fontenc}
\smartqed  

\RequirePackage{graphicx}
\RequirePackage{mathptmx}      
\RequirePackage[numbers,sort&compress]{natbib}
\RequirePackage[colorlinks,citecolor=blue,urlcolor=blue,linkcolor=blue]{hyperref}

\RequirePackage{color}
\RequirePackage{float}
\RequirePackage{setspace}
\RequirePackage{latexsym,amssymb,alltt}
\RequirePackage{fancyhdr}
\RequirePackage{supertabular}

\RequirePackage{epsf}
\RequirePackage{units}
\RequirePackage{lineno}
\RequirePackage{amsmath}

\newcommand{\numu}{$\nu_{\mu}$ }
\newcommand{\nue}{$\nu_{e}$ }
\newcommand{\nutau}{$\nu_{\tau}$ }
\newcommand{\dg}{$^\circ$}
\newcommand{\truee}{{\textsc{Truee}}}

\journalname{Eur. Phys. J. C}

\begin{document}

\title{Development of a General Analysis and Unfolding Scheme and its Application to Measure the Energy Spectrum of Atmospheric Neutrinos with IceCube
}


\onecolumn

\author{IceCube Collaboration: M. G. Aartsen\thanksref{Adelaide}
\and M.~Ackermann\thanksref{Zeuthen}
\and J.~Adams\thanksref{Christchurch}
\and J.~A.~Aguilar\thanksref{Geneva}
\and M.~Ahlers\thanksref{MadisonPAC}
\and M.~Ahrens\thanksref{StockholmOKC}
\and D.~Altmann\thanksref{Erlangen}
\and T.~Anderson\thanksref{PennPhys}
\and C.~Arguelles\thanksref{MadisonPAC}
\and T.~C.~Arlen\thanksref{PennPhys}
% ...
}
\authorrunning{IceCube Collaboration}
\thankstext{a}{present address Department of Physics and Astronomy, Michigan State University, 567 Wilson Road, East Lansing, MI 48824, USA}
\thankstext{b}{Earthquake Research Institute, University of Tokyo, Bunkyo, Tokyo 113-0032, Japan}
\thankstext{c}{NASA Goddard Space Flight Center, Greenbelt, MD 20771, USA}
\institute{
% ...
School of Chemistry \& Physics, University of Adelaide, Adelaide SA, 5005 Australia \label{Adelaide}
% ...
\and Dept. of Physics and Astronomy, University of Canterbury, Private Bag 4800, Christchurch, New Zealand \label{Christchurch}
% ...
\and Erlangen Centre for Astroparticle Physics, Friedrich-Alexander-Universit\"at Erlangen-N\"urnberg, D-91058 Erlangen, Germany \label{Erlangen}
\and D\'epartement de physique nucl\'eaire et corpusculaire, Universit\'e de Gen\`eve, CH-1211 Gen\`eve, Switzerland \label{Geneva}
% ...
\and Dept. of Physics and Wisconsin IceCube Particle Astrophysics Center, University of Wisconsin, Madison, WI 53706, USA \label{MadisonPAC}
% ...
\and Oskar Klein Centre and Dept. of Physics, Stockholm University, SE-10691 Stockholm, Sweden \label{StockholmOKC}
% ...
\and Dept. of Physics, Pennsylvania State University, University Park, PA 16802, USA \label{PennPhys}
% ...
\and DESY, D-15735 Zeuthen, Germany \label{Zeuthen}
}

\date{Received: date / Accepted: date}

\maketitle

\twocolumn

\begin{abstract}
We present the development and application of a generic analysis scheme for the measurement of neutrino spectra with the IceCube detector. This scheme is based on regularized unfolding, preceded by an event selection which uses a Minimum Redundancy Maximum Relevance algorithm to select the relevant variables and a Random Forest for the classification of events. The analysis has been developed using IceCube data from the 59-string configuration of the detector. 
27,771 neutrino candidates were detected in 346 days of livetime. A 
rejection of 99.9999\% of the atmospheric muon background is achieved. The energy spectrum of
the atmospheric neutrino flux is obtained using the TRUEE unfolding program. The unfolded spectrum of
atmospheric muon neutrinos covers an energy range from 100 GeV to 1 PeV. Compared to the previous measurement
using the detector in the 40-string configuration, the analysis presented here, extends the upper end of
the atmospheric neutrino spectrum by more than a factor of two, reaching an energy region that has not
been previously accessed by spectral measurements.

\end{abstract}

\section{Introduction}

Measuring the energy spectrum of atmospheric muon neutrinos is particularly challenging due to its steeply falling behavior. As neutrinos cannot be detected directly, their flux is measured through the detection of neutrino-induced muons. However, atmospheric muons produced in extended air showers when a cosmic ray interacts with a nucleus in the Earth's atmosphere constitute a natural background to atmospheric neutrino searches. In a detector like IceCube~\cite{IceCubeFirstYear}, the majority of this atmospheric muon background can be rejected by the selection of upward going tracks. Remaining background events consist of originally downward-going muons falsely reconstructed as upward going. Thus, an effective selection of events is required. 

Furthermore, the energy of the neutrino cannot be accessed directly, but needs to be inferred from energy dependent observables. These challenges demand a sophisticated data analysis chain, considering both the separation of signal and background events and the reconstruction of the spectrum by using unfolding techniques. 

This paper describes a novel analysis approach aimed at measuring the
atmospheric muon-neutrino spectrum. We use experimental data taken with IceCube
in the 59-string configuration. The analysis consists of an event selection
based on a data pre-processing using quality cuts on a few selected variables,
followed by a machine learning algorithm for final event selection.

In a machine learning algorithm events are classified according to their properties. Rules for this classification are automatically derived from a set of events for which the class is known, e.g. simulated events. The induction of classification rules is generally referred to as training.

All analysis steps were carefully validated and are based on well established methods
from Computer Science and Statistics. This approach was found to outperform previous
measurements~\cite{IC40Atmospheric} with respect to background rejection and signal efficiency.
We then present the first application of the new unfolding program TRUEE~\cite{TrueePaper} on
IceCube data. This analysis procedure proved capable of producing a neutrino energy spectrum
from $100~\unit{GeV}$ to $1~\unit{PeV}$.

The paper is organized as follows: In section~\ref{IceCube} we describe the IceCube detector. Section~\ref{neutrinos} summarizes the basic physics of atmospheric neutrinos. The machine learning algorithms used for event selection, their validation and their application to IceCube data are covered in section~\ref{events}. An enhanced unfolding algorithm and its application in an atmospheric neutrino analysis are presented in section~\ref{unfolding}. In section~\ref{AngularUnfolding} the spectrum is unfolded for two different zenith bins. A comparison of the results to previous measurements is given in section~\ref{comparison}. A summary of the results concludes the paper (section~\ref{Summary}).

\section{IceCube}\label{IceCube}

IceCube is a cubic-kilometer neutrino detector located at the geographic South Pole.
Neutrinos are detected through the Cherenkov light emitted by secondary particles produced in
neutrino-nucleon interactions in or around the detector. The detector consists of an
array of digital optical modules (DOMs) mounted on 86 cables (or strings). The strings are
arranged in an hexagon with typical horizontal spacing of $125~\unit{m}$, and hold 60 DOMs each.
The vertical separation between DOMs is 17~m and they are deployed at depths between
$1450~\unit{m}$ and $2450~\unit{m}$. Eight strings at the center of the array were deployed with a
distance of about 70~m and vertical DOM distance of $7~\unit{m}$. This denser configuration is part
of the DeepCore detector~\cite{DeYoung2009}.
Each DOM consists of a $25~\unit{cm}$ Hamamatsu R7081-02 Photo-multiplier Tube (PMT) and a suite of
electronics board assemblies contained within a spherical glass pressure housing of $35.6~\unit{cm}$
diameter. High accuracy and a wide dynamic range can be achieved by the DOMs by internally
digitizing and time-stamping the photonic signals. Packaged digitized data is then
transmitted to the surface. Each DOM can operate as a complete and autonomous
data acquisition system~\cite{IceCubeFirstYear,IceCubeDAQ}. IceCube was successfully completed in December 2010.

IceTop stations are located on the top of the strings, forming an air-shower array with a nominal grid spacing matching the $125~\unit{m}$ of the in-ice part of the detector. Each station consists of two tanks equipped with downward facing DOMs with their lower hemisphere embedded in the ice. Two DOMs are deployed per tank for redundancy and flexibility~\cite{IceCubeFirstYear}.

The Cherenkov light emitted by muons produced in neutrino interactions can be used to reconstruct the muon trajectory. Since at high energies (TeV or above) the direction of the muon deviates only marginally from the direction of the neutrino, the direction of the incoming neutrino can be reconstructed as well. The pointing resolution of IceCube was found to be 0.7\dg~in a moon shadow analysis using TeV cosmic rays~\cite{MoonShadow2013}. 

There are two primary detection channels in IceCube, the first one being \emph{track-like} events originating from charged current (CC) \numu interactions of the form:
\begin{equation}
\nu_{\mu}+N\longrightarrow \mu + X, 
\label{eq:CCInteraction}
\end{equation}
where $N$ represents a nucleon and $X$ denotes the rest of the particles produced in the interaction. The second channel are \emph{cascade-like} events produced in CC interactions of \nue and \nutau and in neutral current (NC) interactions of all neutrino flavors. Only \numu CC interactions are relevant for the atmospheric neutrino analysis presented in this paper.

Data for this analysis were taken between May 2009 and May 2010, when the detector consisted of 59 strings. This  configuration is referred to as IceCube-59. The analysis is based on a preselection of events which is provided to the analyzers by the IceCube collaboration.

\section{Atmospheric Neutrinos}\label{neutrinos}

Although primarily designed for the detection of high-energy neutrinos from astrophysical sources, IceCube can also be used for investigating the atmospheric neutrino spectrum over several orders of magnitude in energy. Despite the fact that the atmospheric \numu spectrum has been measured by various experiments including Frejus~\cite{FrejusAtmospheric}, AMANDA~\cite{AmandaAtmospheric}, ANTARES~\cite{ANTARES} and IceCube in the 40-string configuration~\cite{IC40Atmospheric}, the flux, especially at high energies, is still subject to rather large uncertainties~\cite{Fedynitch2012}. 

The flux of atmospheric muon neutrinos is dominated by neutrinos originating from the decay of pions and kaons, produced in extended air showers, up to energies of $E_{\nu}\approx500~\unit{TeV}$~\cite{AmandaAtmospheric} (conventional atmospheric neutrino flux). Due to their relatively long lifetime, pions and kaons lose part of their energy prior to decaying. As the flux of cosmic rays follows a power law, the atmospheric neutrino spectrum is also expected to follow a power law, which is one power steeper (asymptotically $\frac{d\Phi}{dE}\propto E^{-3.7}$) compared to the spectrum of primary cosmic rays~\cite{IC40Atmospheric}. 

However, despite the isotropic distribution of cosmic rays, the flux of
conventional atmospheric neutrinos is a function of the zenith angle,
since horizontally travelling mesons have a much higher probability to decay
before losing energy in collisions~\cite{Gaisser1990}. 
This results in a harder neutrino
spectrum of horizontal events compared to vertical events.

At energies exceeding $500~\unit{TeV}$, neutrinos from the decay of charmed mesons, so called prompt neutrinos, are expected to contribute notably to the spectrum. Since neutrinos from the decay of charmed mesons have not been conclusively detected, the exact threshold depends strongly on the underlying model. Due to their short lifetime ($t_{\textrm{life}}\approx10^{-12}~\unit{s}$~\cite{PDG}), these mesons decay before interacting and follow the initial spectrum of cosmic rays more closely, therefore causing a flattening of the overall neutrino flux~\cite{IC40Atmospheric,AmandaAtmospheric}. 

A detailed measurement of the conventional and prompt atmospheric neutrino spectrum is made difficult by its steeply falling characteristic and the finite energy resolution of neutrino energy reconstruction. We have developed an analysis technique making use of machine learning processes to select a sample of neutrino candidates with high purity.

\section{Event Selection}\label{events}

The signature of atmospheric muons entering the detector from above is similar to the event pattern of a neutrino-induced muon. Both signatures can be distinguished by their reconstructed track parameters and quality measures, which form an $n$-dimensional parameter space. Selecting events from this parameter space can be achieved by making good use of machine learning algorithms.

Selecting only upward going tracks can remove a large fraction of the atmospheric muon background. A certain fraction of muon events, however, is falsely reconstructed as upward going. This type of event still occurs 1,000 times more frequently than neutrino-induced events. As mis-reconstruc\-ted muons are significantly harder to reject, a multi-faceted event selection needs to be carried out to obtain a highly pure sample of neutrino candidates.

The event selection presented here consists of several consecutive steps: Initially, two simple cuts are applied to reduce the event sample to a manageable size. As a second step, variables to be used as input for the learner are selected using an automated variable selection.  As IceCube runs multiple reconstruction algorithms on each interesting event, there are hundreds of variables that are potential inputs to the classification algorithm. 
We use an automated variable selection process to select the variables that have the most power for separating signal and background events. Data preprocessing, variable selection and performance of the classification algorithm were thoroughly validated in cross validations, where the average performance over many splits in disjoint training and test data is obtained.

\subsection{Data Preprocessing}

The preprocessing consisted of a cut on the LineFit velocity ($v_{\textrm{LineFit}}>0.19~\unit{c}$) and a cut on the reconstructed zenith angle ($\theta>88^{\circ}$)\footnote{A reconstructed zenith angle of 0\dg corresponds to an event entering the detector from above (the South), whereas a reconstructed zenith angle of 180\dg corresponds to an event entering the detector from below (the North).}. 

The LineFit algorithm reconstructs a track on the basis of the position, $\vec{r}_{i}$, and hit times, $t_{i}$, of all DOMs with a hit in the event. The geometry of the Cherenkov cone as well as the optical properties of the medium are ignored, and the method assumes that the photons propagate along a 1-dimensional line with constant speed, $\vec{v}_{\textrm{LineFit}}$. Minimizing the following $X^{2}$:
\begin{equation}
X^{2} = \sum_{i}^{N} (\vec{r}_{i} - \vec{r}_{\textrm{LineFit}} - \vec{v}_{\textrm{LineFit}} \cdot t_{i})^{2},
\label{eq:Chi2}
\end{equation}
one obtains the fit parameters, $\vec{v}_{\textrm{LineFit}}$ and $\vec{r}_{\textrm{LineFit}}$, where $i$ runs over the DOMs with a hit in the event. Cascade-like events will produce a spherical light pattern from which small values of $\left| v_{\textrm{LineFit}}\right|$ are reconstructed. As long muon tracks of high quality are required for a reliable reconstruction of the energy spectrum, a cut on $\left | v_{\textrm{LineFit}}\right |$ can be utilized to select such events.

The zenith-angle cut is aimed at reducing the contamination of atmospheric muons entering the detector at angles $\theta<90^{\circ}$. Choosing a cut at $\theta=88^{\circ}$ rather than at $\theta=90^{\circ}$ aims at a slight extension of the field of view in order to detect high energy neutrinos from above the horizon. Muons approaching the detector at angles between $\theta=88^{\circ}$ and $\theta=90^{\circ}$, are very likely to range out before reaching the detector. 

Both cuts were optimized simultaneously with respect to background rejection and signal efficiency.
The application of the two cuts yielded a background rejection of 91.4\% at a signal efficiency of 57.1\%.

\subsection{Automated Variable Selection}

The quality of an automated, machine learning-based, event selection largely depends on the set of variables used (in machine learning these are generally referred to as "features" or "attributes"). In this analysis the variables considered as input for the learner were the reconstructed properties of the events and different measures of the quality of the reconstruction. As not all variables are equally well suited for the event selection, and since using all available variables would result in an unreasonably large consumption of computing resources, a representation in fewer dimensions needs to be found. In general, a manual selection based on knowledge about the detector and the classification problem at hand will result in a good set of variables for training the classification algorithm. It will, however, not necessarily result in the \textit{best} set of variables. In the event selection presented in this paper, we therefore used the Minimum Redundancy Maximum Relevance (MRMR) algorithm~\cite{DingAndPeng} for the selection of variables. 

Within MRMR the relevance of a set of variables is computed from an $F$-test, whereas its redundancy $V$ can be obtained from the following equation~\cite{DingAndPeng}:
\begin{equation}
V=\dfrac{1}{\left | F \right |^{2}}\sum_{i,j}\left | c(x_{i},x_{j}) \right |,
\label{eq:Redundancy}
\end{equation}
where $F$ represents a set of variables. To compute the similarity between two variables $x_{i}$ and $x_{j}$ the absolute value $\left | c(x_{i},x_{j}) \right |$ of Pearson's correlation coefficient is used. As a final selection criterion the quotient $Q$ between relevance and redundancy is computed. The variable set, which maximizes $Q$ is returned. MRMR is particularly useful when certain quantities (e.g. zenith angle) are obtained from a number of different reconstruction algorithms. For futher details on MRMR we refer to ref.~\cite{DingAndPeng}.

As variable selections are in general carried out on a limited number of events, their performance might be influenced by statistical fluctuations within those subsets. The average performance given by the cross validation is a valid output. However, one might want to additionally inspect the stability of the variable selection. The stability expresses the variance over different cross validation splits. Two stability measures, Jaccard index and Kuncheva's index~\cite{Kuncheva2007} were used to determine the stability of the MRMR variable selection. They express the ratio between the data splits returning the same variables and the number of variables returned by all splits. The basic equation for the Jaccard index is:

\begin{equation}
J=\frac{| F_{i}\cap F_{j} |}{| F_{i} \cup F_{j} |},
\label{eq:JaccardIndex}
\end{equation}
where $F_{i}$ and $F_{j}$ represent two subsets of variables, selected on two disjoint sets of events drawn at random from the same distribution. 

A similar stability measure is Kuncheva's index, defined as:

\begin{equation}
I_{C}(F_{i},F_{j})=\frac{rn-k^{2}}{k(n-k)}.
\label{eq:KunchevasIndex}
\end{equation}
In equation~\eqref{eq:KunchevasIndex} the parameter $k$ represents the size of the subset, whereas $r=| A \cap B |$ represents the cardinality of the intersection. The total number of variables available is denoted by $n$. 

\begin{figure} 
\begin{minipage}{\columnwidth}
\centering
\includegraphics[width=\columnwidth]{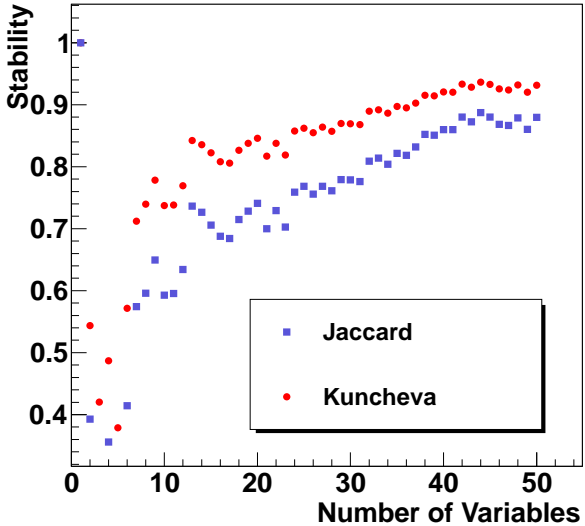}
\end{minipage}
\caption{Stability of the MRMR variable selection as a function of the number of variables considered. The Jaccard and Kuncheva's indices were used as stability measures. One finds that both stability measures increase with the number of variables considered. 
As both measures are well above 0.7, indicating a stable selection, if 25 or more variables are selected, MRMR can be considered stable in case this threshold is exceeded.}
\label{fig:MRMR_Stability}
\end{figure}

The stability of the variable selection was tested with respect to the number of variables selected. To perform this test the number of variables was increased stepwise by one variable in the range between one and 50 variables. For each number of variables the MRMR variable selection was restarted and repeated 10 times on 10 disjoint subsets of events. The overall stability $\bar{S}$ as depicted in Fig.~\ref{fig:MRMR_Stability} is defined as the  average of the indices $I$ for all combinations of those feature selections~\cite{Schowe/Morik/2011b}:
\begin{equation}
\bar{S}=\dfrac{2}{l^{2}-l}\sum_{i=1}^{l}\sum_{j=i+1}^{l}I(F_{i},F_{j}),
\label{eq:OverallStability}
\end{equation}
where $l$ is the total number of feature selections for a specific number of variables. The quantity $I$ in eq.~\ref{eq:OverallStability} represents the Jaccard index or Kuncheva's index, respectively. In total 10,000 events were used for the calculation of the indices.

The stability measures presented in Eqs.~\ref{eq:JaccardIndex} and~\ref{eq:KunchevasIndex} can take values between 0 and 1. In general a selection is considered stable if the indices are close to 1 and considered unstable if the indices are close to 0. 
Figure~\ref{fig:MRMR_Stability} depicts the stability of the MRMR variable selection as a function of the number of selected variables. The stability of the variable selection is found to increase with the number of variables selected. It is further observed that both stability measures are well above 0.7, in case the number of selected variables exceeds 25. 

Twenty-five variables were selected as this number represents a reasonable trade-off between variable selection stability and the anticipated resource consumption of the learner. Moreover, the separation power of the remaining variables was found to be close to zero.

Attributes found to yield large separation power in this analysis are zenith angles, the length of the track obtained from direct photons and the number of direct photons detected in various time windows. Photons are referred to as \textit{direct} when their arrival time at the DOM agrees with that expected for unscattered cherenkov photons~\cite{Ahrens2004}.

\subsection{Performance of the Random Forest}

In general, the evaluation of the performance of a classification algorithm consists of the two important steps of training and testing the algorithm. From the machine learning point of view the event selection can be formalized in terms of a classification task with the classes {\em signal} (atmospheric neutrinos) and {\em background} (atmospheric muons). 

A Random Forest~\cite{Breiman2001}, which utilizes an ensemble of simple decision trees, was chosen as the machine learning algorithm because ensemble algorithms are well known for their robustness and stability. In general trees can be interpreted easily and performed well in previous IceCube analyses~\cite{IC40Atmospheric}. Moreover, a study by Bock et al. has shown that Random Forests outperform other classification algorithms~\cite{BockMagic}. Training and testing were carried out in a standard five-fold cross-validation. 

Within the cross validation 70,000 simulated neutrino events and 750,000 simulated background events were used. In a cross validation the labeled events are split into five disjoint subsets of events. In every iteration one of the disjoint sets is used to test the performance of the Random Forest, whereas the remaining sets are used for training. Thus, 14,000 neutrino events and 150,000 background events were available for testing in every iteration in the five-fold cross validation used in this analysis. Accordingly, 56,000 neutrino events and 600,000 background events per iteration were available for training. The neutrino events were generated by the IceCube neutrino-generator (NuGen). Background events were simulated according to the Polygonato model~\cite{Polygonato2003} using CORSIKA~\cite{Corsika}.

The 25 variables selected by the MRMR algorithm were used for the training of the forest. In order to improve the overall performance of the event selection three additional parameters were created and added according to the findings in~\cite{IC40Atmospheric}. The first variable added is the absolute difference between the zenith angle obtained from a simple LineFit and the reconstructed zenith angle obtained from a multi-photo-electron (MPE) fit. As a second variable the difference between the log-likelihood obtained from a Bayesian fit and a single-photo-electron (SPE) fit was added. The third variable added was the log-likelihood derived from an MPE-fit, divided by the number of hit DOMs. For details on the individual fit algorithms we refer to~\cite{Ahrens2004}.

Within the forest, every event is labeled as signal or background according to its attributes by every tree. The final output score is then computed by averaging over the classifications of the individual trees in the forest.

The ratio of signal and background events used for training the forest was varied systematically. These tests yielded that the signal-to-background ratio available for training did not result in an optimal performance of the learner. Within the tests it was found that very good results in terms of signal efficiency and background rejection can be obtained using 27,000 simulated signal- and 27,000 simulated background events for the training of the forest. Furthermore, a reasonable trade-off between signal efficiency and background rejection could be achieved using this setting. In order to provide the learner with this number of events a simple sampling operation was carried out inside the cross validation. Within this sampling 27,000 simulated neutrino events and 27,000 simulated background events were drawn at random. Helping the learning algorithm by using balanced training and test sets does 
not imply that the application of the learned function works only on balanced class distributions. Empirically, we have observed that the decision function obtained from balanced samples can be successfully applied to extremely biased samples.  As the sampling only concerned the training of the Random Forest, the number of events available for testing remained unchanged. 

The neutrino events used in the training process were simulated according to an $E^{-2}$ flux. Using an $E^{-2}$ flux instead of an atmospheric neutrino flux will provide the learner with enough events also at high energies. This is required in order to obtain a reliable classification over the entire energy range. Although this flux deviates from an atmospheric neutrino flux it can still be used for the training of the forest as the classification is achieved on an event-by-event basis. Therefore, once a certain event pattern is {\it memorized} as neutrino-like by the forest, events with similar patterns will always be labelled as signal, independent of the underlying energy distribution. Furthermore, the result achieved using a decision tree depends only weakly on the underlying distribution used for training. After classification every event was re-weighted according to an atmospheric flux in order to obtain a prediction of the neutrino rate. 

In general, the performance of a Random Forest is found to increase with the number of trees. However, the larger the number of trees, the larger the computational cost for training and testing (CPU time and memory). It was found that 500 trees provided a reasonable tradeoff between the performance of the classification algorithm and the computational cost. Therefore, the forest was trained and validated using 500 trees.

\begin{figure} 
\begin{minipage}{\columnwidth}
\centering
\includegraphics[width=\columnwidth]{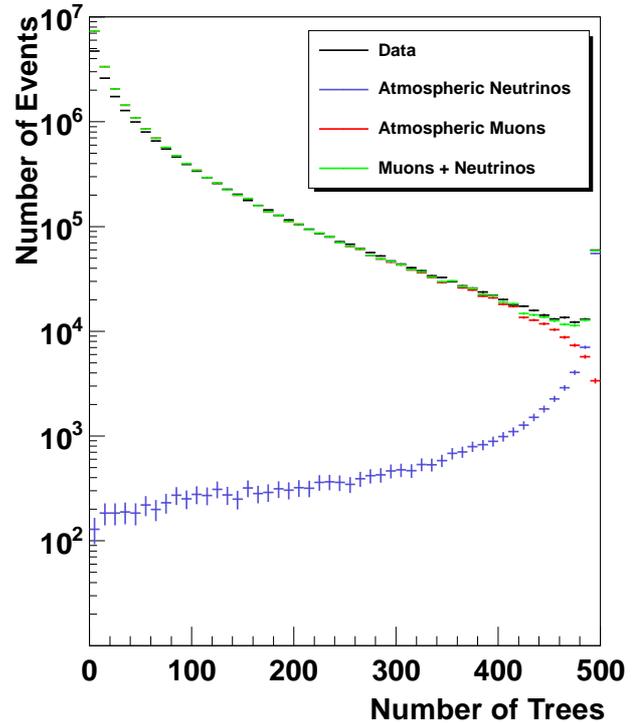}
\end{minipage}
\caption{Number of Trees classifying an event as signal. Atmospheric neutrinos are depicted in blue, whereas atmospheric muons are shown in red. Experimental data is shown in black, whereas the sum of simulated signal and background events is depicted in green. The sum of simulated signal events and background events is found to agree well with the distribution of experimental data, indicating a stable performance of the Random Forest.}
\label{fig:ConfidenceScaled}
\end{figure}
\begin{figure} 
\begin{minipage}{\columnwidth}
\centering
\includegraphics[width=\columnwidth]{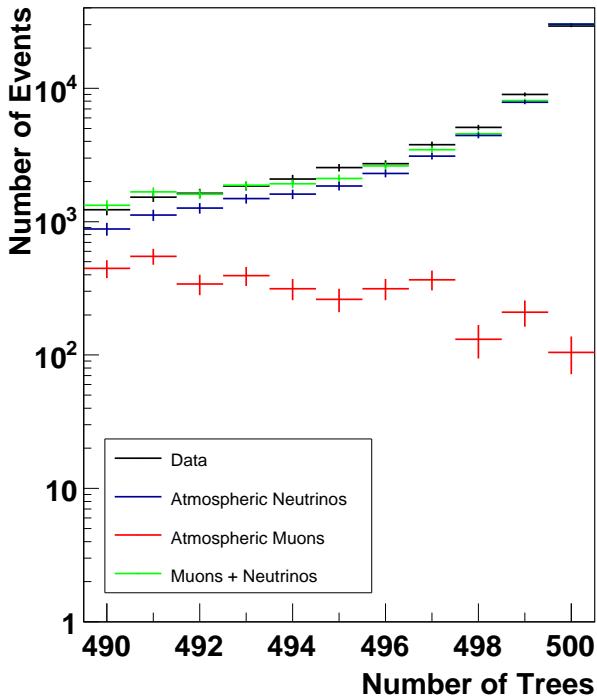}
\end{minipage}
\caption{Same as Fig.~\ref{fig:ConfidenceScaled}, zoom into the region where the final selection cut is considered.} 
\label{fig:ConfidenceZoomed}
\end{figure}
The output scores of the Random Forest for simulated events and experimental data are shown in Fig.~\ref{fig:ConfidenceScaled} and Fig.~\ref{fig:ConfidenceZoomed}. Figure~\ref{fig:ConfidenceZoomed} focuses on the region between 490 and 500 trees, whereas the entire output range of the Random Forest is depicted in Fig.~\ref{fig:ConfidenceScaled}. The well matching distributions of experimental data and simulated events indicate a stable performance of the forest. The rather poor agreement of simulated events and experimental data for $n_{\textrm{trees}}<100$ originates from poorly reconstructed muons of low energy. 

Unfolding the energy distribution of the neutrino sample requires an extremely strict rejection of atmospheric muons. This is due to the fact that only a small number of events is found to populate the highest energy bins. Therefore, a single high energy muon might cause a flattening of the unfolded spectrum at high energies and thus mimic a prompt or astrophysical flux of neutrinos. We chose a very strict cut of $n_{\textrm{trees}}=500$, thus selecting only events that were classified as signal by every tree in the forest. 

The {\em statistical uncertainty} of the event selection, which is introduced due to statistical fluctuations in the training and test sets, was estimated from the cross validation results. The statistical uncertainty can be calculated from the signal efficiency and background rejection of the individual iterations. A statistical uncertainty of 1.6\% was estimated for the expected number of neutrino candidates, which indicates a stable and reliable performance of the forest.

The {\em systematic uncertainty} of the event selection was estimated by applying the forest to simulated events produced with different DOM efficiencies and a different modeling of the ice. For this purpose the efficiencies of all DOMs were either increased or decreased by 10\% from their nominal values. The modeling of the ice was taken into account by using the SPICE Mie ice model~\cite{SpiceMie} instead of its predecessor SPICE-1. It was found that the uncertainty of the event selection due to the ice model is on the order of 5\%, whereas the uncertainty due to the DOM efficiency was estimated to be 18\%. Combining both values one finds that the total systematic uncertainty of the event selection is 19\%. 

After verifying the performance of the Random Forest the final model was trained using 27,000 simulated neutrino events and 27,000 simulated background events. The events for each class were drawn at random from the total sample of available simulated events. 

The application of the entire event selection chain on the full set of IceCube-59 data yielded 27,771 neutrino candidates in 346 days of detector live-time ($\approx80$ neutrino candidates per day). The number of remaining atmospheric muons was estimated to be $114\pm103$. The purity of the final neutrino event sample was estimated to be $(99.59^{+0.36}_{-0.37})$\%. No events with a zenith angle $\theta<90^{\circ}$ were observed in the sample after the application of the Random Forest.

The number of events surviving the two preselection cuts on the zenith angle and the LineFit velocity is $15.3\times10^{6}$. This corresponds to an estimated background rejection of $91.4$\% at a signal efficiency of $57.1$\%.

Comparing the total number of neutrino candidates at final level an increase of 62\% is observed with respect to~\cite{IC40Atmospheric}, which used IceCube in the 40-string configuration. Taking into account the larger volume of the detector (59 compared to 40 strings) and the increased trigger rate, the event selection method presented in this paper succeeds in an increase of 8\% in the number of neutrino candidates compared to the event selection presented in~\cite{IC40Atmospheric}. The relative contamination of the sample with atmospheric muons was found to be of the same size as in~\cite{IC40Atmospheric}.

In the event selection, which is the basis for the subsequent unfolding of the \numu energy spectrum, a signal efficiency of 18.2\% was achieved at a background rejection of  $99.9999$\%, which corresponds to a reduction of the contamination of the event sample with atmospheric muons by six orders of magnitude. Both signal efficiency and background rejection were computed for events with $\theta_{\textrm{Zenith}}\geq88$\dg, with respect to the starting level of the analysis and for neutrino energies between $E_{\nu}=100~\unit{GeV}$ and $E_{\nu}=1~\unit{PeV}$.

All event selection steps regarding machine learning, preprocessing, and validation were carried out using the \textsc{RapidMiner}~\cite{RapidMiner} machine learning environment.

\section{Spectrum Unfolding}\label{unfolding}

As the neutrino energy spectrum cannot be accessed directly, it needs to be inferred from the reconstructed energy of the muons. This task is generally referred to as an inverse, or ill-posed, problem and described by the Fredholm integral equation of first kind~\cite{TrueePaper}: 

\begin{equation}
g(y)=\int^{a}_{b}A(y,E)\,f(E)\,dE.
\end{equation}
For the discrete case this transforms to:

\begin{equation}
\vec{g}(y)=\underline{A}(y,E)\vec{f}(E),
\label{eq:Fredholm}
\end{equation}
where $\vec{f}(E)$ is the sought energy distribution and the measured energy dependent distribution is given as $\vec{g}(y)$. The matrix $\underline{A}(y,E)$ represents the response matrix of the detector, which accounts for the physics of neutrino interactions in or near the detector as well as for the propagation of the muon. 

Several approaches to the solution of inverse problems exist. The unfolding program \textsc{Truee}~\cite{TrueePaper}, which is an extension of the $\mathcal{RUN}$~\cite{RunManual} algorithm, was used for unfolding in this analysis. The stability of the unfolding as well as the results obtained on experimental data are addressed in the following. 

\subsection{Unfolding Input}

The spectrum is unfolded in ten logarithmic energy bins between $100~\unit{GeV}$ and $1~\unit{PeV}$. Three variables (track length, number of hit DOMs, number of direct photons) were used as input for the unfolding.  Direct hits have not suffered scattering in the ice from their emission point to the DOM and therefore keep precise timing information, which is essential for an accurate track reconstruction. For the unfolding only direct hits from a time window ranging from $-15,\unit{ns}$ to $75,\unit{ns}$ have been used. An estimate of the track length inside the detector is obtained by projecting all DOMs that recorded direct photons onto the reconstructed track. 

\begin{figure} 
\begin{minipage}{\columnwidth}
\centering
\includegraphics[width=\columnwidth]{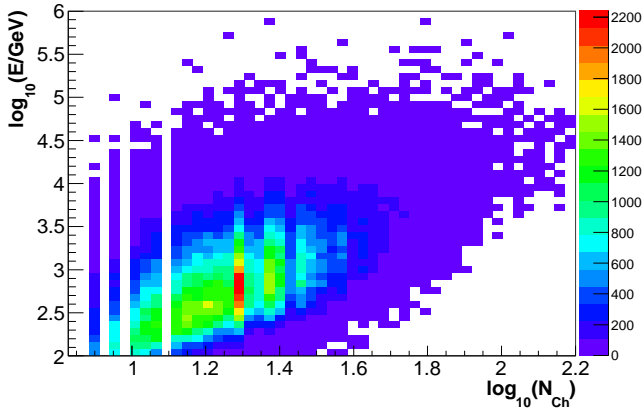}
\end{minipage}
   \caption{Neutrino energy $E$ vs. the number of hit DOMs ($N_{\textrm{Ch}}$) for the simulated events used for the determination of the response matrix.}
   \label{fig:E_vs_NCh}
\end{figure}

\begin{figure} 
\begin{minipage}{\columnwidth}
\centering
\includegraphics[width=\columnwidth]{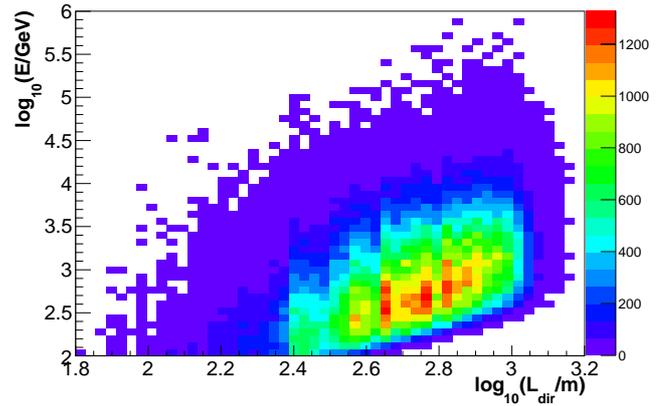}
\end{minipage}
   \caption{Neutrino energy $E$ vs. the estimated track length inside the detector $L_{\textrm{dir}}$ for the simulated events used for the determination of the response matrix.}
   \label{fig:E_vs_LDirC}
\end{figure}

\begin{figure} 
\begin{minipage}{\columnwidth}
\centering
\includegraphics[width=\columnwidth]{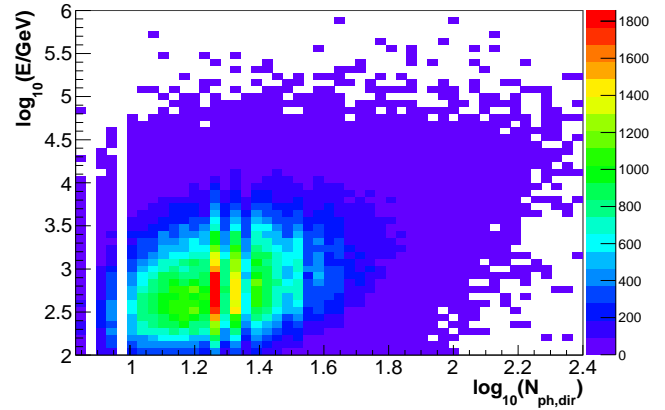}
\end{minipage}
    \caption{Neutrino energy $E$ vs. the number of direct photons $N_{\textrm{ph,dir}}$ for the simulated events used for the determination of the response matrix.}
    \label{fig:E_vs_NDirC}
\end{figure}

The energy dependence of the three input variables for simulated events is depicted in Figs.~\ref{fig:E_vs_NCh}-\ref{fig:E_vs_NDirC}. Good correlation with energy was found for all three observables. A sample of 300,000 simulated neutrino events was used for the determination of the response matrix. This number corresponds to approximately ten times the livetime of IceCube in the 59-string configuration. The sample was obtained by sampling events according to their atmospheric weights. The energy distribution of simulated events thus, matches the one of an atmospheric neutrino spectrum. 
\subsection{Verification}
\begin{figure} 
\begin{minipage}{\columnwidth}
\centering
\includegraphics[width=\columnwidth]{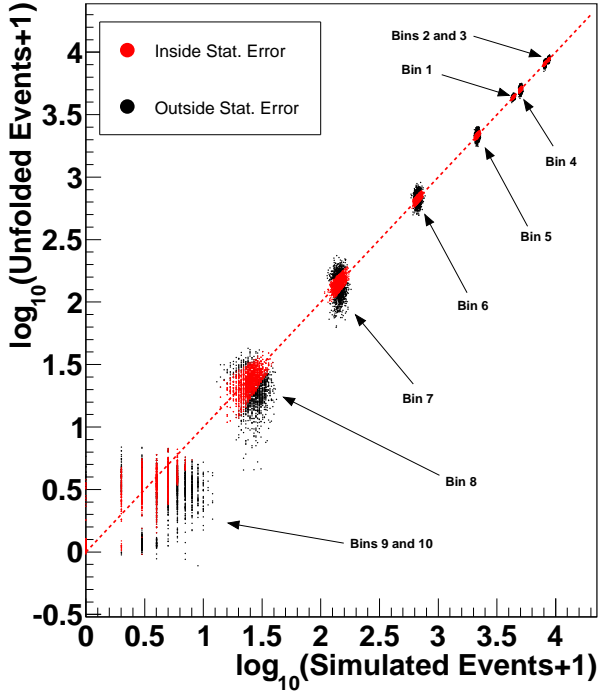}
\end{minipage}
\caption{Results of 500 unfoldings for all bins. The x-axis depicts the number of simulated events, whereas the number of unfolded events is shown on the y-axis. Unfoldings where the difference between the true number of events in a certain bin and the unfolding result for that bin lies within the statistical uncertainty returned by \truee~are shown in red.
Unfoldings where this is not the case are depicted in black. The energy spectrum of the simulated events corresponds to an atmospheric spectrum. In general, we find that the number of unfolded events is highly correlated with the true number of events in a certain unfolding. The individual populations observed in the plot, correspond to the individual energy bins of the unfolded distribution.} 
\label{fig:PullModeResult}  
\end{figure}
The verification of the unfolding result consists of two different tests. The first test is based on multiple unfoldings of a specified number of simulated events, which are drawn at random. This kind of test can be accessed via \truee~\cite{TrueePaper}. The second test is based on re-weighting simulated events according to the unfolded spectrum of atmospheric $\nu_{\mu}$. Both tests were successfully carried out and are individually addressed in the following. 

The result of the first test is shown in Fig.~\ref{fig:PullModeResult}. Within this test a fraction of simulated events is drawn at random. For every bin the unfolding result is then compared to the number of injected events in that bin. For the analysis reported here 500 test unfoldings were carried out. The number of injected events from the Monte Carlo distribution is depicted on the x-axis of Fig~\ref{fig:PullModeResult} and the number of unfolded events is shown on the y-axis.

The individual populations observed in the figure correspond to the individual energy bins of the final unfolding result. The line-like structures observed for small event numbers are due to the fact that only integers are possible as event number for the true MC distributions, whereas real numbers can be returned as the unfolding result for the individual 
bins. 

The rather large deviation between the unfolding result and the number of injected events obtained for the highest energy bins is a result of the steeply falling spectrum of atmospheric neutrinos and the applied bootstrapping procedure. Due to the small number of expected events in the last bin, either 0 or 1 events are drawn randomly from the true distribution. Two or more events are only drawn in rather rare cases. Based on the response matrix, which accounts for the limited statistics in the highest energy bins by using ten times more events compared to experimental data, only a fraction of an event is reconstructed for the highest energy bin. As the statistical uncertainties derived in \textsc{Truee} fail to account for the difference between the predicted bin content and the number of injected events, large deviations are observed. This further implies that an overestimation is obtained in case no events are present in the last bin on experimental data. As soon as one event is present 
in this bin, 
an 
underestimation is 
observed. 

Within \truee the statistical uncertainties are computed as the square root of the diagonal elements of the covariance matrix. This test can therefore be used to validate the statistical uncertainties returned by the algorithm. The unfolding result is compared to the underlying distribution of events. If the difference between the unfolding result and the true value is covered by the statistical uncertainty returned by \truee, the statistical uncertainties are estimated correctly. For cases where the statistical uncertainty fails to cover this difference the statistical uncertainty is scaled up. For the analysis presented here an underestimation of the number of injected events is observed for the 9th and 10th bin, respectively. This underestimation is not covered by the statistical uncertainty, which is thus scaled up by a factor of 1.9 for the 9th, and a factor of 6.3 for the 10th bin.

In a second test, simulated events are re-weighted according to the unfolding result (see Fig.~\ref{fig:TRUEE_Output} of section~\ref{sec:UnfoldingResult}). For a successful unfolding, data and simulated events are expected to agree after re-weighting. This test was carried out for the three variables used as input for the unfolding but also for two additional energy dependent observables (energy loss per unit length $dE/dX$ and total charge $Q_{\textrm{tot}}$).

\begin{figure} 
\begin{minipage}{\columnwidth}
\centering
\includegraphics[width=\columnwidth]{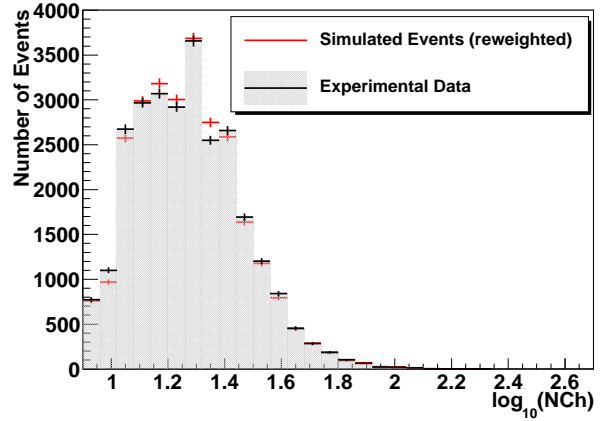}
\end{minipage}
 \caption{Simulated events (red) re-weighted to the unfolding result (Fig.~\ref{fig:TRUEE_Output}) compared to real data (black) for the number of hit DOMs $N_{\textrm{Ch}}$.}
 \label{fig:CheckUnfolding_NCh}
\end{figure}

\begin{figure} 
\begin{minipage}{\columnwidth}
\centering
\includegraphics[width=\columnwidth]{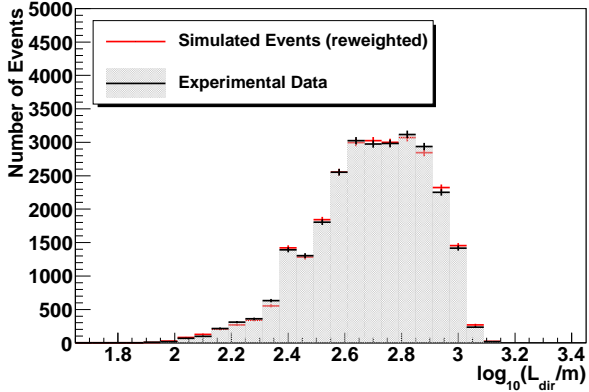}
\end{minipage}
  \caption{Simulated events (red) re-weighted to the unfolding result (Fig.~\ref{fig:TRUEE_Output}) compared to real data (black) for the estimated track length inside the detector $L_{\textrm{dir}}$.}
  \label{fig:CheckUnfolding_LDirC}
\end{figure}

\begin{figure} 
\begin{minipage}{\columnwidth}
\centering
\includegraphics[width=\columnwidth]{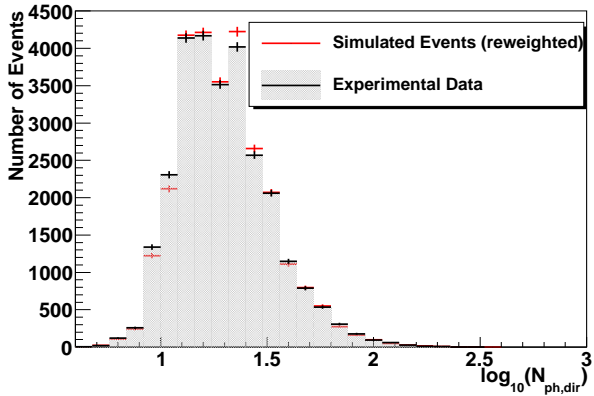}
\end{minipage}
  \caption{Simulated events (red) re-weighted to the unfolding result (Fig.~\ref{fig:TRUEE_Output}) compared to real data (black) for the number of direct photons $N_{\textrm{ph,dir}}$.}
  \label{fig:CheckUnfolding_NDirC}
\end{figure}

\begin{figure} 
\begin{minipage}{\columnwidth}
\centering
\includegraphics[width=\columnwidth]{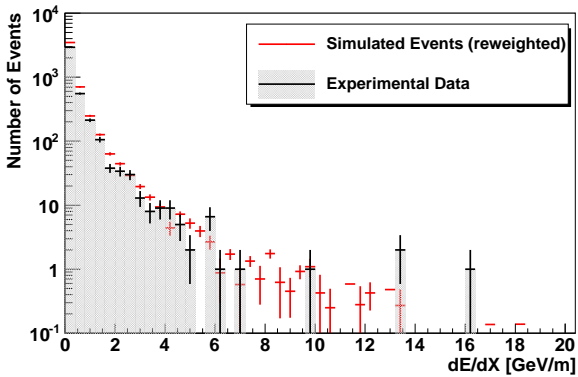}
\end{minipage}
  \caption{Simulated events (red) re-weighted to the unfolding result (Fig.~\ref{fig:TRUEE_Output}) compared to real data (black) for the energy loss per unit length $dE/dX$.}
  \label{fig:CheckUnfolding_dEdX}
\end{figure}

\begin{figure} 
\begin{minipage}{\columnwidth}
\centering
\includegraphics[width=\columnwidth]{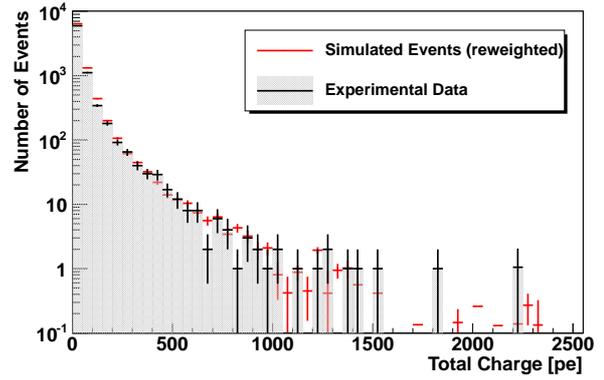}
\end{minipage}
  \caption{Simulated events (red) re-weighted to the unfolding result (Fig.~\ref{fig:TRUEE_Output}) compared to real data (black) for the total charge collected in an event $Q_{\textrm{tot}}$.}
  \label{fig:CheckUnfolding_QTot}
\end{figure}

The outcome of the re-weighting is depicted in Figs.~\ref{fig:CheckUnfolding_NCh} to~\ref{fig:CheckUnfolding_QTot}. A good agreement between data and the re-weighted simulation is observed over the entire range of the individual parameters.
\subsection{Estimation of systematic uncertainties}

As the unfolding result is obtained by using a response matrix determined from Monte Carlo simulation, the properties of the simulation will affect the unfolding result. In order to determine the effect of different simulation settings on the spectrum of atmospheric neutrinos, additional unfoldings were carried out using different sets of simulated events for the determination of the response matrix. For each simulation set used for the estimation of systematic uncertainties one property was changed with respect to the default simulation set.

The setting for the efficiency of the DOMs was varied by $\pm$10\% with respect to the nominal value. Within this simulation the efficiency of all DOMs was simultaneously increased or decreased, respectively. A shift of $\pm10$\% with respect to the nominal value is slightly larger than the $7.7$\% cited in~\cite{DOMCalibration} and is thus a bit more conservative.

Further systematic tests were carried out by using simulated events generated with a $\pm5$\% increased and decreased pair production cross section, respectively. The value of $\pm5\%$ was chosen to be slightly more conservative than the theoretical uncertainty cited in~\cite{Kokoulin1999}. The modeling of the ice was varied as well, by using the SPICE Mie ice model~\cite{SpiceMie} instead of its predecessor SPICE-1.

The response matrices obtained for the individual systematic sets of data were then applied to real data in order to estimate the size of the systematic uncertainties. Prior to the application on real data, however, every setting was checked using the multiple unfoldings in \truee. No indications for instabilities were observed for any of the systematic tests.

Thus, five additional unfolding results were obtained on real data. The difference between the unfolding result obtained using the standard Monte Carlo sets and the systematic Monte Carlo sets were computed bin-wise and for every setting. The final uncertainties were calculated by adding the obtained differences in quadrature. This procedure further offers the advantage that all systematic uncertainties are derived on experimental data.

For energies up to $1\,\unit{TeV}$ the total systematic uncertainty is dominated by the uncertainty arising from the modeling of the ice. For energies above $1\,\unit{TeV}$ the uncertainties due to the DOM efficiencies and the modeling of the ice were found to be of approximately the same size. A more precise modeling of the ice and a better understanding of the DOM efficiency, is therfore likely to reduce the systematic uncertainties of future measurements.

\subsection{Unfolding Result}
\label{sec:UnfoldingResult}
\begin{figure} 
\begin{minipage}{\columnwidth}
\centering
\includegraphics[width=\columnwidth]{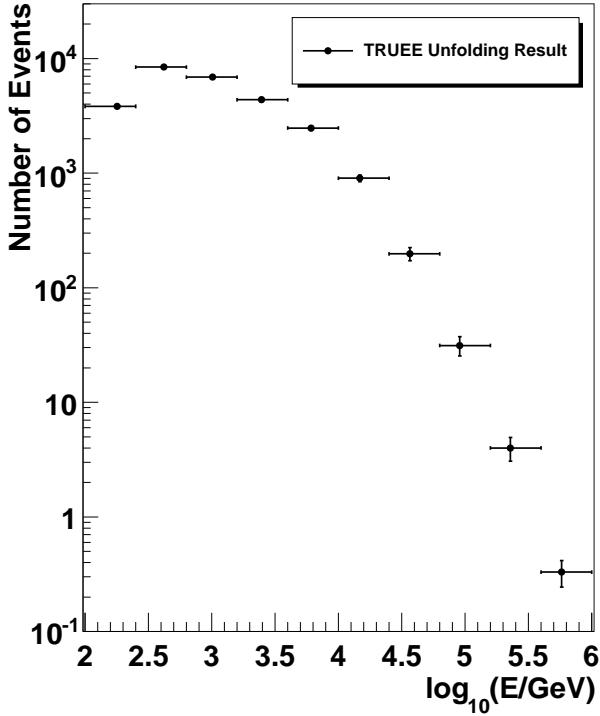}
\end{minipage}
  \caption{Number of unfolded events per bin as returned by \truee.}
  \label{fig:TRUEE_Output}
\end{figure}

\begin{figure} 
\begin{minipage}{\columnwidth}
\centering
\includegraphics[width=\columnwidth]{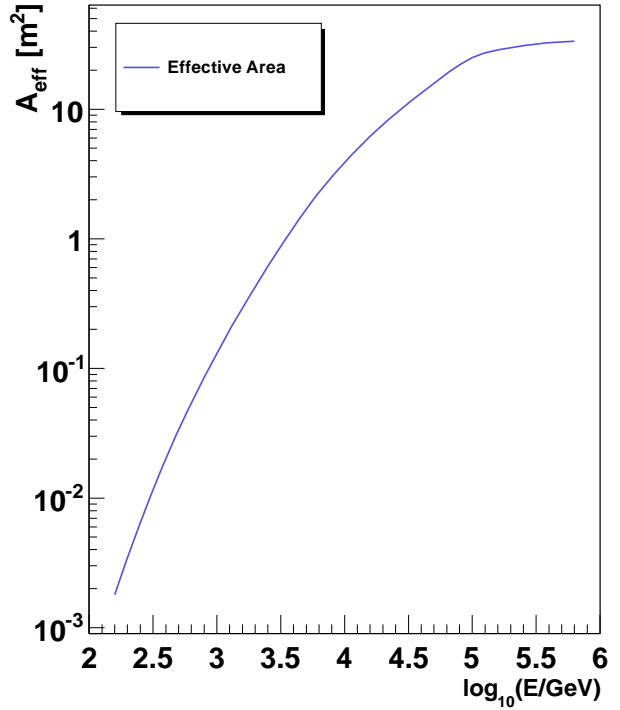}
\end{minipage}
 \caption{Neutrino effective area for the analysis presented in this paper.} 
\label{fig:EffectiveArea} 
\end{figure}

The number of unfolded events as returned by \truee~is depicted in Fig.~\ref{fig:TRUEE_Output}. The energies of the bins were obtained as the mean of the distribution of simulated atmospheric neutrino events for every bin. This result can now be converted into a flux of atmospheric neutrinos by utilizing the effective area $A_{\textrm{eff}}$ and the livetime of the detector as well as the solid angle. The effective area for this analysis is shown in Fig.~\ref{fig:EffectiveArea}.
\begin{figure} 
\begin{minipage}{\columnwidth}
\centering
\includegraphics[width=\columnwidth]{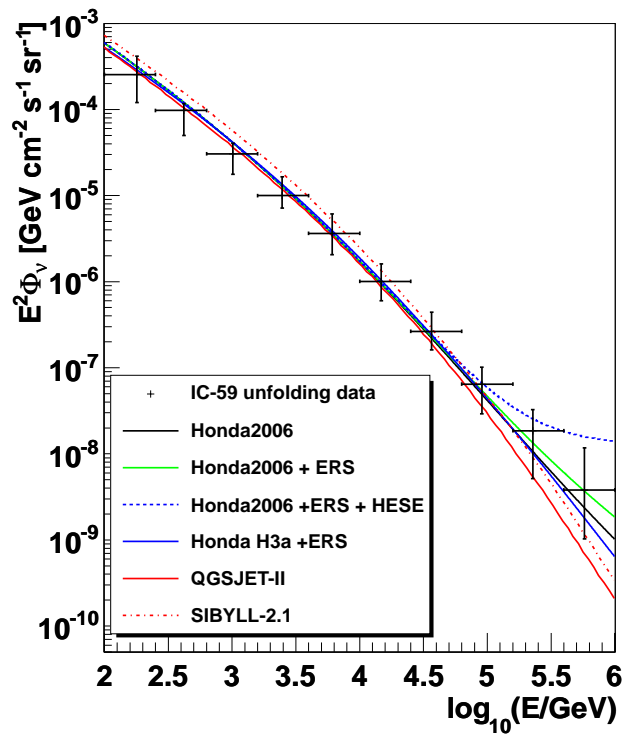}
\end{minipage}
 \caption{Acceptance corrected flux of atmospheric neutrinos from 100~GeV to 1~PeV, compared to several theoretical models (please see the text for more details on the individual models).} 
\label{fig:ComparisonPlot} 
\end{figure}
Figure~\ref{fig:ComparisonPlot} shows the acceptance-corrected and zenith-averaged flux of atmospheric neutrinos obtained with IceCube in the 59-string configuration of the detector. The spectrum covers the energy range from $100~\unit{GeV}$ to $1~\unit{PeV}$. Six theoretical model expectations are shown for comparison. The model by Honda et al.~\cite{Honda2007} (Honda2006), extrapolated to higher energies is shown as a solid black line. The Honda2006 model only models the conventional atmospheric neutrino flux. The model by Honda et al. together with a model of the prompt component by Enberg et al.~\cite{Sarcevic2008} (ERS) is shown as a solid green line. The recent best fit to an astrophysical flux obtained in the IceCube high energy starting event analysis (HESE)~\cite{IceCubeScience} are included as a third component in the blue dashed line. An additional modeling of the knee of the cosmic ray flux is included in the model labeled Honda H3a + ERS (solid blue line). Atmospheric neutrino flux predictions 
obtained 
from ANFlux~\cite{Fedynitch2012} using QGSJET-II~\cite{QGSJET} and SIBYLL-2.1~\cite{SIBYLL} as hadronic interaction models are shown as a solid red line and a red dashed-dotted line respectively. 

Compared to the IceCube-40 result the systematic uncertainties of the spectrum were reduced, especially at low and intermediate energies. The decreased error bars are due to a better understanding of systematic effects in IceCube. 
Due to the relatively large systematic uncertainties at high energies, no statement can be made about a possible contribution of neutrinos from the decay of charmed mesons. Furthermore, no statement about a possible contribution of neutrinos from astrophysical sources can be made in this analysis. 

In general, a good agreement between the unfolded flux and the models is observed. Deviations of $3.2~\sigma$ and $2.6~\sigma$ are observed between the unfolded distribution and the theoretical model obtained using SIBYLL-2.1 as a hadronic interaction model, for the second ($E_{\nu}=418~\unit{GeV}$) and third bin ($E_{\nu}=1013~\unit{GeV}$), respectively. However, a correlation of the systematic uncertainties of these two bins should be noted.

The acceptance-corrected flux of atmospheric neutrinos as well as the relative uncertainties are summarized in Tab.~\ref{tab:FinalResult}.
\begin{table}
\centering
\caption{Bin-wise summary of the acceptance-corrected unfolding result, which corresponds to the differential flux of atmospheric neutrinos, scaled by $E^{2}$ and given in $\unit{GeV~cm^{-2}~sr^{-1}~s^{-1}}$.}
\label{tab:FinalResult}  
\begin{tabular*}{\columnwidth}{@{\extracolsep{\fill}}llll@{}}
\hline
$\log_{10}(E/\unit{GeV})$ & $E^{2}\Phi$ & $\sigma_{rel.}^{stat.}$ [\%] & $\sigma_{rel.}^{syst.}$ [\%] \\
\hline
2.25 & $2.54\times10^{-4}$ & $\pm2.5$ & $^{+63}_{-53}$ \\
 2.62 & $0.97\times10^{-4}$ & $\pm2.3$ & $^{+19}_{-49}$ \\
 3.01 & $3.06\times10^{-5}$ & $\pm3.2$ & $^{+32}_{-42}$\\
 3.39 & $1.00\times10^{-5}$ & $\pm4.4$ & $^{+65}_{-28}$\\
 3.78 & $3.64\times10^{-6}$ & $\pm4.5$ & $^{+69}_{-43}$\\
 4.17 & $1.01\times10^{-6}$ & $\pm6.7$ & $^{+60}_{-40}$\\
 4.56 & $2.65\times10^{-7}$ & $\pm13.1$ & $^{+66}_{-37}$\\
 4.96 & $6.44\times10^{-8}$ & $\pm19.0$ & $^{+54}_{-52}$\\
 5.36 & $1.85\times10^{-8}$ & $^{+45.8}_{-23.5}$ & $^{+61}_{-68}$\\
 5.76 & $3.81\times10^{-9}$ & $^{+163}_{-26.0}$ & $^{+130}_{-68}$\\
\hline
\end{tabular*}
\end{table}
\section{Unfolding of Different Angular Regions}
\label{AngularUnfolding}
\begin{figure} 
\begin{minipage}{\columnwidth}
\centering
\includegraphics[width=\columnwidth]{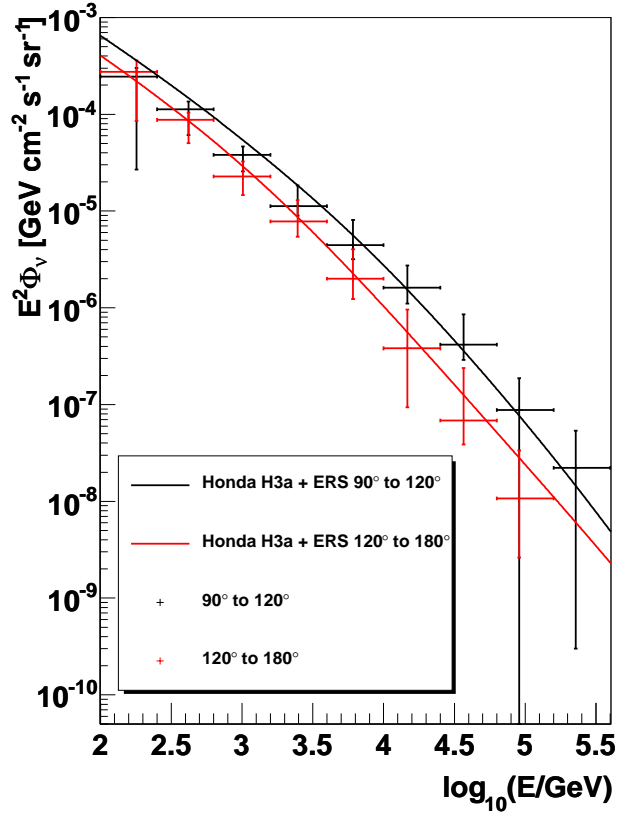}
\end{minipage}
 \caption{Unfolded atmospheric neutrino flux for the energy range from $100~\unit{GeV}$ to $316~\unit{TeV}$ and for two different zenith bands. Events with a reconstructed zenith angle from $90^{\circ}$ to $120^{\circ}$ are depicted in black, whereas events with a reconstructed zenith angle from $120^{\circ}$ to $180^{\circ}$ are shown in red. The Honda H3a+ERS model is shown for comparison. Compared to the neutrino spectrum obtained for the full angular range, a smaller range in energy is covered, which is due to the smaller statistics of the two unfolded samples.}
\label{fig:AngularResults}
\end{figure}
In order to study the dependence of the atmospheric neutrino flux on the zenith angle, additional unfoldings were carried out dividing the data into two separate sets according to the reconstructed zenith angle. The first zenith band contains events with a reconstructed zenith angle between 90$^{\circ}$ and 120$^{\circ}$, whereas events with reconstructed zenith angles between $120^{\circ}$ and $180^{\circ}$ were used for the second zenith band. Using the 500 unfoldings of simulated events selected randomly it was found that no changes in the unfolding settings were required in order to unfold the two different angular regions. The same input parameters as for the unfolding of the full angular range were used and the systematic uncertainties were estimated in the same way as described above. Because of the smaller statistics the unfolding was not extended as high in energy as for the full sample. The upper end of the spectrum extends to $E_{\nu}=316~\unit{TeV}$ for events with a reconstructed zenith angle between 90\dg~and 120\dg. 
An 
upper end of $E_{\nu}=158~\unit{TeV}$ is reached for events with a reconstructed zenith angle between 120\dg~and 180\dg.
 
The result of unfolding the two different angular regions is depicted in Fig.~\ref{fig:AngularResults}. The flux obtained for the zenith band from $90^{\circ}$ to $120^{\circ}$ is depicted in black, whereas the flux obtained for the zenith band from $120^{\circ}$ to $180^{\circ}$ is shown in red. The Honda2006 model, accounting for a different modeling of the knee plus using the ERS model for the prompt component of the atmospheric flux, is shown for both angular regions for comparison. 

In general, a good agreement between the unfolded distribution and the theoretical model is observed. The unfolding results for the two angular bins are summarized in Tab.~\ref{tab:FinalResult90to120} and Tab.~\ref{tab:FinalResult120to180}
\begin{table}
\centering
\caption{Bin-wise summary of the acceptance-corrected unfolding result for zenith angles between 90\dg~and 120\dg, which corresponds to the differential flux of atmospheric neutrinos, scaled by $E^{2}$ and given in $\unit{GeV~cm^{-2}~sr^{-1}~s^{-1}}$.}
\label{tab:FinalResult90to120}  
\begin{tabular*}{\columnwidth}{@{\extracolsep{\fill}}llll@{}}
\hline
$\log_{10}(E/\unit{GeV})$ & $E^{2}\Phi$ & $\sigma_{rel.}^{stat.}$ [\%] & $\sigma_{rel.}^{syst.}$ [\%] \\
\hline
2.25 & $2.45\times10^{-4}$ & $\pm4.3$ & $^{+23}_{-89}$ \\
 2.62 & $1.13\times10^{-4}$ & $\pm3.2$ & $^{+20}_{-46}$ \\
 3.01 & $3.80\times10^{-5}$ & $\pm3.9$ & $^{+22}_{-32}$\\
 3.39 & $1.12\times10^{-5}$ & $\pm5.5$ & $^{+63}_{-19}$\\
 3.78 & $4.45\times10^{-6}$ & $\pm5.8$ & $^{+82}_{-28}$\\
 4.17 & $1.61\times10^{-6}$ & $\pm7.2$ & $^{+70}_{-31}$\\
 4.56 & $4.15\times10^{-7}$ & $\pm13.9$ & $^{+105}_{-27}$\\
 4.96 & $8.76\times10^{-8}$ & $\pm22.2$ & $^{+112}_{-115}$\\
 5.36 & $2.22\times10^{-8}$ & $^{+58.2}_{-29.1}$ & $^{+129}_{-94}$\\
\hline
\end{tabular*}
\end{table}
\begin{table}
\centering
\caption{Bin-wise summary of the acceptance-corrected unfolding result for zenith angles between 120\dg~and 180\dg, which corresponds to the differential flux of atmospheric neutrinos, scaled by $E^{2}$ and given in $\unit{GeV~cm^{-2}~sr^{-1}~s^{-1}}$.}
\label{tab:FinalResult120to180}
\begin{tabular*}{\columnwidth}{@{\extracolsep{\fill}}llll@{}}
\hline
$\log_{10}(E/\unit{GeV})$ & $E^{2}\Phi$ & $\sigma_{rel.}^{stat.}$ [\%] & $\sigma_{rel.}^{syst.}$ [\%] \\
\hline
2.25 & $2.75\times10^{-4}$ & $\pm3.1$ & $^{+31}_{-69}$ \\
 2.62 & $0.87\times10^{-4}$ & $\pm3.4$ & $^{+19}_{-42}$ \\
 3.01 & $2.28\times10^{-5}$ & $\pm4.7$ & $^{+43}_{-35}$\\
 3.39 & $7.81\times10^{-6}$ & $\pm5.6$ & $^{+65}_{-30}$\\
 3.78 & $1.99\times10^{-6}$ & $\pm7.3$ & $^{+102 }_{-37}$\\
 4.17 & $3.81\times10^{-7}$ & $\pm17.4$ & $^{+151}_{-73}$\\
 4.56 & $6.84\times10^{-8}$ & $\pm36.5$ & $^{+247}_{-24}$\\
 4.96 & $1.07\times10^{-8}$ & $\pm52.7$ & $^{+207}_{-54}$\\
\hline
\end{tabular*}
\end{table}
\section{Comparison to Previous Experiments}
\label{comparison}
\begin{figure} 
\begin{minipage}{\columnwidth}
\centering
\includegraphics[width=\columnwidth]{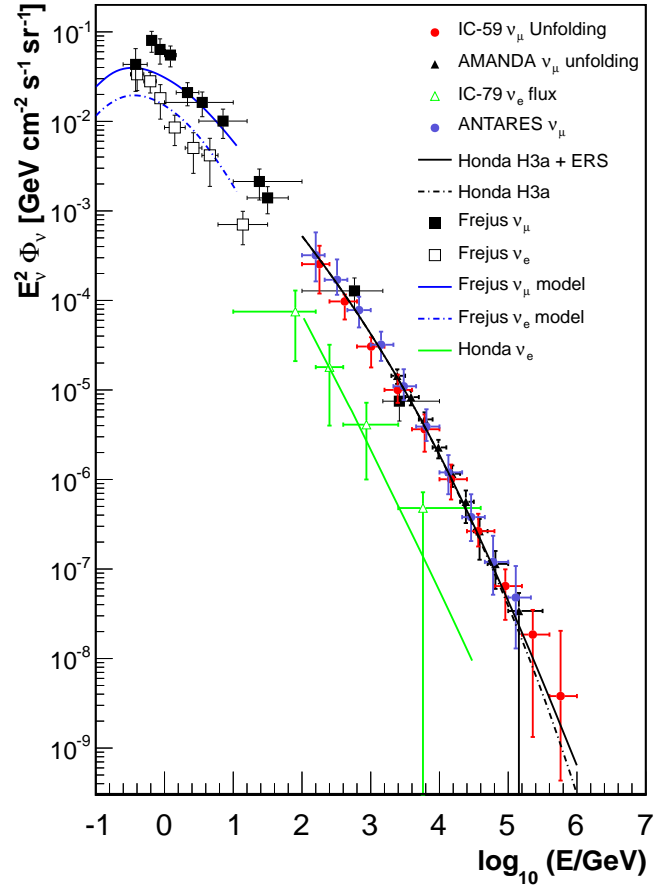}
\end{minipage}
\caption{Comparison of the unfolding result obtained using IceCube in the 59-string configuration to previous experiments. At the low energy end of the spectrum the results of the Frejus experiment~\cite{FrejusAtmospheric} are depicted as black squares for \numu, whereas the Frejus results for $\nu_{e}$ are shown as hollow squares. The unfolding results obtained with the AMANDA experiment~\cite{AmandaAtmospheric} are shown as black triangles. Results from the ANTARES neutrino telescope~\cite{ANTARES} are depicted in blue. The $\nu_{e}$ spectrum obtained using IceCube in the 79 string configuration~\cite{IC79AtmosphericCascades} is shown as green triangles. The results of the analysis presented here are shown as red circles. Theoretical models are shown for comparison.}
\label{fig:AllResults}   
\end{figure}
Figure~\ref{fig:AllResults} shows the results of the measurement presented in this paper, depicted as red circles, in the wider context of measurements obtained with previous experiments. 
We find that the results derived in this measurement are in good agreement with both the theoretical models and previous measurements of the atmospheric \numu flux. Comparing our results to the spectrum obtained using the AMANDA detector we find that the measurement extends to energies that are larger by almost an order of magnitude. The two measurements are found to agree well within their estimated systematic uncertainties. Due to the different energy thresholds, the IceCube and Frejus spectra overlap only between $100~\unit{GeV}$ and $1~\unit{TeV}$. Both measurements agree within their error bars. Comparing the measurement presented in this paper to the results obtained with the ANTARES neutrino telescope~\cite{ANTARES} we find that both measurements are fully compatible within their systematic uncertainties. A gap in experimental data points exists at energies between $30~\unit{GeV}$ and $300~\unit{GeV}$. Within this energy region neutrino oscillations become important and, thus, the spectrum becomes more complicated. This gap can most likely be closed by 
utilizing the full capabilities of IceCube DeepCore, which has an 
energy 
threshold of $10~\unit{GeV}$~\cite{DeepCoreDesign}. The measurement presented here did not benefit from the more densely instrumented DeepCore strings, as only one such string had been deployed at the time of the measurement. 

\section{Summary}
\label{Summary}
In this paper we presented the measurement of the atmospheric \numu flux obtained using IceCube in the 59-string configuration. The unfolded spectrum of atmospheric muon neutrinos covers an energy range from $100~\unit{GeV}$ to $1~\unit{PeV}$, thus covering four orders of magnitude in energy. Compared to the previous measurement of the atmospheric \numu flux, which utilized the detector in the 40-string configuration, the analysis presented here extended the upper end of the atmospheric neutrino spectrum by more than a factor of two. 

This increase in the accessible energy was achieved by using a dedicated event selection procedure, which utilized state of the art algorithms from the field of machine learning and data mining. Using a Random Forest preceded by an Minimum Redundancy Maximum Relevance variable selection we were able to reject $99.9999$\% of the incoming background events. At this background rejection 27,771 atmospheric neutrino candidates were detected in $346~\unit{days}$ of IceCube-59. This corresponds to $80.3$ neutrino events per day, which is a significant improvement over the 49.3 neutrino events per day reported in~\cite{IC40Atmospheric}. 
The purity of the final neutrino sample was estimated to $(99.59^{+0.36}_{-0.37})$\%. Taking into account the excellent agreement between expectations derived on the basis of simulated events and results obtained on experimental data (see Fig.~\ref{fig:ConfidenceScaled}) we find that the combination of a Random Forest and an MRMR can be applied to real life problems, delivering excellent results in terms of both background rejection and signal efficiency. 

An energy spectrum of the atmospheric \numu was obtained using the new unfolding software \truee. The unfolding result was validated using a bootstrapping procedure implemented in \textsc{Truee}. A test using multiple unfoldings of simulated neutrino events selected at random yielded a very good agreement between the unfolding result and the true distribution of events, thus validating the overall stability of the unfolding process. Comparing the unfolding results to theoretical models, one finds that no statement on a possible contribution of a prompt and/or astrophysical component to the overall flux of atmospheric neutrinos can be made, due to the relatively large uncertainties at high energies. 

Additional years of measurements with IceCube in the 79-string and in the 86-string configurations are likely to confirm the results from~\cite{IceCubeScience} in spectral measurements. It is further expected that the systematic uncertainties will decrease due to a better understanding of systematic effects and due to the homogeneous shape of the detector.

In summary we find that the data analysis chain presented in this paper yields highly stable results for both event selection and the reconstruction of the spectrum. The entire analysis procedure can therefore be applied to all other sets of IceCube data with only minor changes. The analysis chain is especially well suited for measurements of the atmospheric neutrino flux, where future analyzers only have to account for the different detector geometry. 

\begin{acknowledgements}

We acknowledge the support from the following agencies:
U.S. National Science Foundation-Office of Polar Programs,
U.S. National Science Foundation-Physics Division,
University of Wisconsin Alumni Research Foundation,
the Grid Laboratory Of Wisconsin (GLOW) grid infrastructure at the University of Wisconsin - Madison, the Open Science Grid (OSG) grid infrastructure;
U.S. Department of Energy, and National Energy Research Scientific Computing Center,
the Louisiana Optical Network Initiative (LONI) grid computing resources;
Natural Sciences and Engineering Research Council of Canada,
WestGrid and Compute/Calcul Canada;
Swedish Research Council,
Swedish Polar Research Secretariat,
Swedish National Infrastructure for Computing (SNIC),
and Knut and Alice Wallenberg Foundation, Sweden;
German Ministry for Education and Research (BMBF),
Deutsche Forschungsgemeinschaft (DFG),
Helmholtz Alliance for Astroparticle Physics (HAP),
Research Department of Plasmas with Complex Interactions (Bochum), Germany;
Fund for Scientific Research (FNRS-FWO),
FWO Odysseus programme,
Flanders Institute to encourage scientific and technological research in industry (IWT),
Belgian Federal Science Policy Office (Belspo);
University of Oxford, United Kingdom;
Marsden Fund, New Zealand;
Australian Research Council;
Japan Society for Promotion of Science (JSPS);
the Swiss National Science Foundation (SNSF), Switzerland;
National Research Foundation of Korea (NRF);
Danish National Research Foundation, Denmark (DNRF)

\end{acknowledgements}

\bibliography{Bibliography}{}
\bibliographystyle{spphys}

\end{document}